\documentclass[aps,twocolumn,showpacs,eqsecnum]{revtex4}

\usepackage{amsmath}
\usepackage{amsfonts}
\usepackage{amssymb}
\usepackage{color}

\definecolor{dyellow}{rgb}{1.,0.8,.0}
\definecolor{myblue}{rgb}{.1,.1,.7}
\definecolor{dcyan}{rgb}{.0,.6,.6}
\definecolor{dmagenta}{rgb}{0.6,0.0,0.6}
\definecolor{brown}{rgb}{0.6,0.2,0.}
\definecolor{darkblue}{rgb}{.0,.0,0.5}
\definecolor{darkred}{rgb}{0.75,0.0,0.0}
\definecolor{orange}{rgb}{1.,.6,.0}
\definecolor{dorange}{rgb}{0.8,.4,.0}
\definecolor{darkgreen}{rgb}{0.0,0.6,0.0}
\definecolor{purple}{rgb}{.4,.0,.4}

\def\red{\color{red}}

\def\bc{\begin{center}}
\def\nno{\nonumber}
\def\ec{\end{center}}
\def\be{\begin{eqnarray}}
\def\ee{\end{eqnarray}}

\newcommand{\omits}[1]{}
\newcommand{\dS}{\textrm{dS}}
\newcommand{\AdS}{\textrm{AdS}}
\newcommand{\FLT}{\textrm{FLTs}}

\newcommand{\HsM}{$H_\theta \subset M_{1,4}/M_{2,3}$}

\newcommand{\M}{$M_{1,3}$}
\newcommand{\BdS}{\textrm{BdS}}
\newcommand{\BAdS}{\textrm{BAdS}}
\newcommand{\Mink}{Mink}

\newcommand{\CG}{$SO(2,4)$}
\newcommand{\cg}{$\mathfrak{so}(2,4)$}

\newcommand{\NsM}{${\cal N}\subset M_{2,4}$}
\newcommand{\CFT}{CFT}
\newcommand{\diag}{\mathrm{diag}}
\newcommand{\vect}[1]{\ensuremath{\boldsymbol{#1}}}

\def\ka{\kappa}
\def\la{\lambda}

\def\d#1#2{\frac{\displaystyle #1}{\displaystyle #2}}

\newcommand\btd{\raise 2pt
\hbox{$\hat\bigtriangledown$}\hskip 1.5pt}
\newcommand\bt{\raise 2pt
\hbox{$\bigtriangledown$}\hskip 1.5pt}

\def\PRD{{\it Phys. Rev.}~{\bf D}}

\def\PLA{{\it Phys. Lett.}~{\bf A}}

\def\CTP{{\it Commun. Theor. Phys. }}

\begin{document}

\title{Triality of conformal extensions of three kinds of special relativity}

\author{Han-Ying Guo}
\affiliation{
    CCAST (World Laboratory), P.O. Box 8730, Beijing 100080, China,}
\affiliation{
    Institute of Theoretical Physics, Chinese Academy of Sciences,
    Beijing 100080, China}

\author{Bin Zhou}
\affiliation{
  Department of Physics, Beijing Normal University, Beijing 100875, China}

\author{Yu Tian}
\affiliation{
    Department of Physics, Beijing Institute of Technology,
    Beijing 100081, China}
\author{Zhan Xu}
\affiliation{
   Physics Department, Tsinghua University, Beijing 100084, China}

\date{February 8, 2007}

\begin{abstract}
The conformal extensions of three kinds of special relativity with
$ISO(1,3)/SO(1,4)/SO(2,3)$ invariance on \Mink/\dS/\AdS~space, respectively,
are realized on an $SO(2,4)/\mathbb{Z}_2$ invariant projective null cone
$[\mathcal{N}]$ as the (projective) boundary of the 5-d \AdS~ space.
The relations among the conformal \Mink/\dS/\AdS~spaces, the motions of
light signals and the conformal field theories  on them can be given. Thus,
there should be a triality for these conformal issues and the conjectured
\AdS/\CFT\ correspondence.
\end{abstract}



\maketitle


\section{Introduction}

It is well known that in Einstein's  special relativity  on the
Minkowski(\Mink) space \M\, with the \Mink\  line-element $ds^2_M$,
massless particles and light signals move {\it in inertia} along
null geodesics, satisfying $ds_M^2=0$, which are invariant under
conformal transformations.  Thus, the symmetry of their motions
should be enlarged from the Poincar\'e group $ISO(1,3)$ to the
conformal group with 15 parameters.

It has been shown that on an almost equal footing with Einstein's special
relativity, there are two other kinds of the de Sitter (\dS)/anti-\dS\ %
(\AdS)-invariant special relativity on
the \dS/\AdS\ space of radius $R$ with Beltrami coordinate atlas,
denoted as the \BdS/\BAdS\ space, with the Beltrami line-element
$ds^2_{B\pm}$, respectively \cite{Lu, LZG, BdS, BdS2, TdS, IWR, NH,
Lu05}. In the \dS/\AdS\ special relativity, massless particles and
light signals move also {\it in inertia} along null straight
world lines with constant coordinate velocity components and
satisfying $ds_{B\pm}^2=0$. These properties are also invariant
under conformal  transformations, respectively. Thus, the symmetry
of these motions should also be enlarged from the \dS/\AdS-group
$SO(1,4)/SO(2,3)$ to the conformal group, respectively.

Under the conformal  transformations, the \Mink/\dS/\AdS\ space
is not closed, often with some events sent to infinity and vice versa,
respectively.  Thus, each of these spaces must be extended so that they are
closed under the conformal  transformations.  It is also known that
\dS\ and \AdS\ are conformally flat. In the viewpoint of the conformal
triality, however, they are also conformally \dS/\AdS,  respectively. In other
words, \textit{locally} \dS, \AdS\ and \Mink\ spaces are conformally
equivalent.  However, globally it is not the case, due to different topologies
of these spaces.

In this paper, we show that in addition to the conformal extension
of Einstein's special relativity on $M_{1,3}$, the conformal
extension of the \dS/\AdS\, special relativity on 4-d \dS/\AdS\ space
can also be realized on the same null cone $\cal N$ modulo the
projective equivalence ``$\sim$" in a $(2+4)$-d \Mink\ space
$M_{2,4}$, respectively.  In order to be distinguished from
$\mathcal{N}$, the resulted 4-d space as a quotient space is denoted
by $[{\cal N}]:=\mathcal{N}/\!\!\sim$. Actually, the conformal
transformations on each space are induced by the $SO(2,4)$
isometries on $M_{2,4}$ and these spaces are mapped to each other by
certain Weyl conformal mappings.  Thus,  the projective null cone
$[{\cal N}]$ and the physics on it are the conformal extension and
compactification of \M/$\dS_4$/$\AdS_4$, underlying the conformal
extension of Poincar\'e/\dS/\AdS-invariant special relativity,
respectively. And the conformal transformations on \M, $\dS_4$ and
$\AdS_4$ can also be equivalently related.  Thus, the conformal
physics, such as the motions of light signals and the conformal
field theories (\CFT s), can be mapped from one space to another.
Here, by the term \CFT, we mainly refer to theories that are
invariant under the 15-parameter conformal transformation groups.
So, there should be a triality of these conformal issues.

Since the projective boundary  of a 5-d \AdS\ space, $\partial(\AdS_5)$,
is just $[\mathcal{N}]$, the conformal $\dS_4$ and $\AdS_4$ can also be
included in $\partial(\AdS_5)$, similar to the conformal \M.  Thus, if the
\AdS/\CFT\ correspondence \cite{adscft} is conjectured, there should
be three versions of \AdS/\CFT\ correspondence \cite{zg}.

This paper is arranged as follows. In Sec. \ref{sect:SRs}, we very
briefly introduce why there are three kinds of special relativity
and show  why their conformal extensions on the conformal
\Mink/\dS/\AdS\ spaces are eventually on the same $[\cal N]$.  In
Sec. III, we study the physics on these conformal extensions and
their relations. We first show the triality of these conformal
extensions via Weyl mappings for any two of them in
\ref{sect:triality}.  Then we focus on the motion of free massless
particles along the null geodesics in  \ref{sect:nullGeod}. We also
consider the relations among the \CFT s on them and
the triad for the \AdS/\CFT-correspondence in  \ref{sect:CFT}.
Finally, we end with a few remarks.

\section{Three kinds of special relativity and their conformal extensions}
\label{sect:SRs}

In this section, we first very briefly introduce why there are two
{other} kinds of special relativity with \dS/\AdS\, invariance,
respectively, in addition to Einstein's special relativity on the
\Mink\ space. Then we show why their conformal extensions on the
conformal \Mink/\dS/\AdS\ spaces are eventually on the same null-cone
modulo projective equivalence in a $(2+4)$-d \Mink\ space.

\subsection{Three kinds of special relativity}
\label{subsect:SRs}

It has been shown that, based on the principle of relativity and the
postulate on invariant universal constants, the \dS/\AdS\ special
relativity \ on the \dS/\AdS\ space with Beltrami coordinate atlas
can be set up \cite{Lu, LZG, BdS, BdS2, TdS, IWR, NH, Lu05}. Thus,
there are three kinds of special relativity on an almost
equal footing. The principle is the same as that in Einstein's
special relativity and the postulate requires that in addition to
the speed of light $c$ there is another invariant universal constant
$R$ as the curvature radius  of the \dS/\AdS\ space, respectively.

The \dS/\AdS\ space with  radius $R$ can be viewed as a 4-d hyperboloid
${H}_\theta$  $(\theta = \pm 1 )$  embedded in $M_{1,4}$/$M_{2,3}$,
respectively:
\be\label{Hpm}
  && H_\theta:  \eta_{\mu\nu}\, \xi^\mu \xi^\nu - \theta\,(\xi^4)^2
  = \eta_{\theta AB}\xi^A\xi^B=-\theta R^2 \lessgtr 0
  \nonumber \\ \\
  && ds^2_{H_\theta}  = \eta_{\mu\nu}\, d\xi^\mu d\xi^\nu -
  \theta\,(d\xi^4)^2=\eta_{\theta AB}d\xi^Ad\xi^B,
  \label{ds2H}
\ee
where $\mu,\nu=0,\cdots,3; ~ A,B= 0,\cdots, 4$. Let us consider a kind of
{\it uniform ``great circular" motions} for a particle with mass $m_R$, defined
by a conserved 5-d angular momentum on \HsM:
\be\label{angular5a}
  \frac{d{\cal L}^{AB}}{ds_{H_\theta}}=0,\quad
  {\cal L}^{AB}:=m_{R}(\xi^A\frac{d\xi^B}{ds_{H_\theta}}
  - \xi^B\frac{d\xi^A}{ds_{H_\theta}}).%
\ee
There is an Einstein-like formula for the particle
\begin{eqnarray}\label{emla}
  { -\frac{1}{2R^2}{\cal L}^{AB}{\cal L}_{AB}=m_{R}^2,}\quad
  {\cal L}_{AB}=\eta_{\theta AC}\eta_{\theta BD}{\cal L}^{CD}.
\end{eqnarray}
For a massless particle or a light signal with $m_R=0$, similar uniform
``circular" motion can also be defined so long as the proper-time
$ds_{H_\theta}$ is replaced by an affine parameter $\lambda_\theta$ and
there is no $m_R$ in the counterpart of ${\cal L}^{AB}$ in (\ref{angular5a}),
respectively. Namely,
\be\label{L5}
  \frac{dL^{AB}}{d\lambda_{\theta}}=0,\quad
  L^{AB}:=\xi^A K^B - \xi^B K^A,
  \quad K^A:=\frac{d\xi^A}{d\lambda_{\theta}}.
  \nonumber \\
\ee
There is also an Einstein-like formula for the particle
\begin{eqnarray}\label{eml5}
  -\frac{1}{2R^2}{ L}^{AB}{ L}_{AB}=0,\quad
  L_{AB}=\eta_{\theta AC}\eta_{\theta BD}{L}^{CD}.
\end{eqnarray}

These properties are symmetrically transformed under the linear transformations
of the \dS/\AdS-group, respectively.

The Beltrami coordinate atlas without antipodal identification on
$dS_4$/$AdS_4$ can be defined patch by patch \cite{BdS, BdS2}.  In the
patch $\xi^4 > 0$ in $H_+$, say, the Beltrami coordinates are:
\begin{equation}\label{x+}
  x_+^\mu = R\,\xi^\mu/\xi^4,\quad \xi^4 > 0.
\end{equation}
In this patch, there are a condition from (\ref{Hpm}) and a Beltrami metric
from (\ref{ds2H}) \cite{Lu, LZG, BdS, BdS2} as follows
\begin{eqnarray}\label{domain}
  && \sigma(x_+) := 1- R^{-2} \eta_{\mu\nu}x_+^\mu x_+^\nu >0,
  \\
  \label{bhl}
  && ds^2_{B+} = \left[\frac{\eta_{\mu\nu}}{\sigma(x_+)}
  +\frac{\eta_{\mu\sigma}\eta_{\nu\rho}x_+^\sigma x_+^\rho }
  {R^2\sigma(x_+)^2}\right] dx_+^\mu dx_+^\nu.
\end{eqnarray}
They  are invariant under the fractional linear transformations with a common
denominator, denoted as the \FLT\,  of the \dS-group
\be\label{G}
  x^\mu\rightarrow \tilde{x}^\mu&=&\pm\sigma^{1/2}(a)\sigma^{-1}(a,x)
     (x^\nu-a^\nu)D_\nu^\mu,\\ \nno
  D_\nu^\mu&=&L_\nu^\mu+ { R^{-2}}
  \eta_{\nu \la}a^\la a^\ka(\sigma(a)+\sigma^{1/2}(a))^{-1}L_\ka^\mu,\\\nno
  L&:=&(L_\nu^\mu)\in SO(1,3),
\ee
which transform a point $A(a^{\mu})$, $\sigma(a^{\mu})>0$ in  the
Beltrami-system $S(x)$ to the origin in the  system $\tilde S(\tilde x)$.

It can be proved that all geodesics of  the Beltrami-metric (\ref{bhl}) are
indeed straight world lines. Actually,  along a timelike or a null geodesic
a massive particle or a light signal moves with {\it constant} coordinate
velocity components, respectively. For the  particle with mass $m_R$ there is
a set of conserved observables:
\be\label{momt4}
  p^\mu=m_{R}\sigma^{-1}(x)\frac{d x^\mu}{ds},
  \quad\frac{dp^\mu}{ds}=0;
  \\ \label{angular4}
  l^{\mu\nu}=x^\mu p^\nu-x^\nu p^\mu, \quad\frac{dl^{\mu\nu}}{ds}=0.
\ee%
 These are the pseudo 4-momentum, pseudo 4-angular-momentum of
the particle, which constitute the conserved 5-d angular
momentum in (\ref{angular5a}). The second equation in (\ref{momt4})
is just the equation for the timelike geodesic of the Beltrami
metric (\ref{bhl}), from which it follows that the second equation
in (\ref{angular4}) is satisfied and its  coordinate velocity
components are constants: \be\label{im}
  \frac{dx^i}{dt}=v^i={\rm const.},~~i=1,2,3.
\ee
For the case of a massless particle or a light signal, it is also the case
for its motion along the null geodesic.  Namely, there is a set of conserved
observables:
\be\label{k4}
  k^\mu=\sigma^{-1}(x)\frac{d x^\mu}{d\lambda_+},
  \quad\frac{dk^\mu}{d\lambda_+}=0;
  \\ \label{Lk4}
  l^{\mu\nu}=x^\mu k^\nu-x^\nu k^\mu, \quad\frac{dl^{\mu\nu}}{d\lambda_+}=0.
\ee%
 These are the pseudo 4-momentum, pseudo 4-angular-momentum of
the massless particle, which constitute the conserved 5-d
angular momentum in (\ref{L5}).

In terms of $p^\mu$ and $l^{\mu\nu}$, the Einstein-like  formula (\ref{emla})
becomes:
\be\label{eml}
  E^2-{p}\,^2- \frac{1}{2R^2}{ l}_{(1,3)}^2=m_{R}^2,\qquad\\
  E^2=m_{R}^2c^4+{p}^2c^2 + \d {c^2} {R^2} j^2 - \d {c^4}{R^2} b^2,
\ee
with energy $E$, momentum $p^i$, $p_i=\delta_{ij}p^j$, ``boost" $b^i$,
$b_i=\delta_{ij}b^j$ and 3-angular momentum $j^i$, $j_i=\delta_{ij} j^j$.
For the massless case, the formula is the same with $m_R=0$.

Thus, there is indeed  a {\it law of inertia} invariant under the
\FLT\, (\ref{G}) in the Beltrami atlas patch by patch for particles
and  light signals. And the principle and the postulate hold and the
\dS-invariant special relativity can further be set up. Further, the
simultaneity, light-cone, horizon and other issues can be well
formulated in the Beltrami atlas patch by patch.

It is important that for  light signals, $ds^2_{B+}=0$. This leads
to the conformal extension of the \BdS\ space invariant under
conformal transformation group with 15 parameters, in analogy with
that of the \Mink\ space.

Similar approach can be applied to the case of \AdS\ space with $\theta=-1$.

It is also clear that under the limit $R\to\infty$, all these issues go back to
the corresponding ones in Einstein's special relativity\ on the \Mink\ space.

\subsection{The conformal extensions on the null cone}

The conformal extensions of \BdS/\BAdS\ space can be realized first via the
hyperboloid $H_\pm$ by introducing a scaling variable $\kappa\neq0$ and a set of
coordinates $\zeta^{\hat A}$, respectively,
\be\label{Liexi+}
  \zeta^\mu := \kappa \xi^\mu, ~
  \zeta^4 := \kappa \xi^4, ~
  \zeta^5:=\kappa R, \quad {\rm for} ~ H_+;\\
  \zeta^\mu := \kappa \xi^\mu, ~
  \zeta^4 := \kappa R, ~
  \zeta^5 := \kappa \xi^4, \quad {\rm for} ~ H_-,
  \label{Liexi-}
\ee
and secondly back to the Beltrami atlas.  Under such a scaling, eq.~(\ref{Hpm})
turns out to be a null cone
\begin{equation}
  \mathcal{N}:~ \eta_{\hat A \hat B} \zeta^{\hat A} \zeta^{\hat B}
  = 0, \quad \eta_{\hat A \hat B} = \diag(J, -1,1),
  \label{LieS0}
\end{equation}
where $J=\diag(1,-1,-1,-1)$, $\Xi := (\zeta^{\hat{A}}) = (\zeta,
\zeta^4, \zeta^5) \neq 0$. Now $H_\pm$ are parts of \NsM{\red .} The
Beltrami -coordinates $x_\pm^\mu$ can be obtained from coordinates
$\zeta^{\hat{A}}$, too.

The null cone $\mathcal{N}$ (\ref{LieS0}) is  $SO(2,4)$-invariant in $M_{2,4}$
and there is the projective equivalence relation ``$\sim$" on $M_{2,4} - \{0\}$:
$\Xi' \sim \Xi$ if and only if there is a number $c\neq 0$ satisfying
$\zeta'^{\hat{A}} = c \, \zeta^{\hat A}$.  The resulted quotient space
$[\cal N] := {\cal N}/\!\!\sim$ is a 4-d submanifold of $\mathbb{R}P^5$,
homeomorphic to $S^1 \times S^3$.  Intuitively,  an equivalence class of
$\Xi \in {\cal N}$ can be viewed as the null straight line passing through both
$\Xi$ and the origin of $M_{2,4}$.  The origin is not included in the
equivalence class, however.  In this sense $[\mathcal{N}]$ consists of all the
null straight lines through the origin.  Thus, an $SO(2,4)$ transformation on
$M_{2,4}$ induces a transformation on $[{\cal N}]$, sending one null straight
line to another.

Thus, the \dS/\AdS-hyperboloid $H_\theta$ can be embedded into the same
$\mathcal{N}$.  When the metric on $M_{2,4}$ is pulled back to $\mathcal{N}$,
it is conformal to $ds^2_{H_\theta}$:
\be\label{ds2C}
  d\chi^2_{\mathcal{N}} : =
  \eta_{\hat{A}\hat{B}}\,d\zeta^{\hat{A}}_{\mathcal{N}}\,
      d\zeta^{\hat{B}}_{\mathcal{N}}
  = \kappa^2 \, ds^2_{H_\theta}.
\ee
In terms of the inhomogeneous projective coordinates or the Beltrami
coordinates, we have
\be\label{ds2CB}
  d\chi^2_{\mathcal{N}} = \kappa^2 \, ds^2_{B \theta}.
\ee
Consequentially, $SO(2,4)$  transformations on (\ref{LieS0}) induce conformal
transformations on $H_\theta$, respectively:
\begin{eqnarray}
  ds'^2_{H_\theta}
  = \rho^2\,ds^2_{H_\theta},\quad
  \rho = \frac{\kappa}{\kappa'}
  = \left\{
    \begin{array}{ll}
      {\zeta^5}/{\zeta'^5}, \quad & \textrm{for } H_+; \\
      {\zeta^4}/{\zeta'^4}, \quad & \textrm{for } H_-.
    \end{array}
  \right .
\end{eqnarray}
Or
\begin{eqnarray}
  ds'^2_{B\theta}
  = \rho^2\,ds^2_{B\theta},\quad
  \rho = \frac{\kappa}{\kappa'}.
\end{eqnarray}

According to eqs.~(\ref{Liexi+}) and (\ref{Liexi-}), $H_\pm$ can be viewed as
the intersection of $\mathcal{N}$ and the hyperplanes $P_+: \zeta^5 = R$ and
$P_-: \zeta^4 = R$, respectively.  Since $H_\theta$ is only part of
$\mathcal{N}$, with $\zeta^5$ (for $H_+$) or $\zeta^4$ (for $H_-$) nonzero,
it is quite possible for an $SO(2,4)$ transformation to send a point in
$H_\theta$, with nonzero $\zeta^5$ or $\zeta^4$, to another one with zero
$\zeta^5$ or $\zeta^4$, and vice versa.  Thus, $H_\theta$ are, in fact, not
closed under the induced conformal transformations.  To be closed, $H_\theta$
must be extended into the whole $[{\cal N}]$.  Thus, $[{\cal N}]$ is the
conformal extension of both \dS\ and \AdS\ spaces.

It is clear that back to the Beltrami atlas, say (\ref{x+}) for the \BdS, as
inhomogeneous projective coordinates, the conformal \BdS/\BAdS-metric follows,
respectively. Thus, the conformal extension of the \dS/\AdS-invariant special
relativity can be set up for massless particles and  light signals,
respectively.

As is well known,  the conformal \Mink\ space can also be obtained from the the
same null cone (see, e.g. \cite{twistor}). To this end, we introduce a set of
new coordinates
\begin{equation}
  \zeta^{\pm} := (\zeta^5\pm\zeta^4)/\sqrt{2}
\end{equation}
and assign coordinates,  with $R$ the same as before,
\be\label{Mcdt}
  x^\mu := R\,\zeta^\mu/\zeta^-, \quad
  x^+ := R\,\zeta^+/\zeta^-
\ee
to those points with $\zeta^- \neq 0$. Then eq.~(\ref{LieS0}) becomes
$x^+ = - \eta_{\mu\nu}\,x^\mu x^\nu/(2R)$, and the metric (\ref{ds2C}) becomes
\be
  d\chi^2_{\cal N} = (\zeta^-/R)^2 \, ds_M^2,&&
  ds_M^2 := \eta_{\mu\nu} \, dx^\mu dx^\nu.
  \label{ihN}
\ee
And, a $SO(2,4)$ transformation induces a conformal transformation on the
\Mink\ space:
\be\label{cmink}
  ds^2_M \rightarrow ds'^2_M = \rho^2\,ds^2_M, \quad
  \rho = \zeta^-/\zeta'^-.
\ee

Similarly, the \Mink\ space \M\ can be regarded as the intersection of
${\cal N}$ and the hyperplane $P_M: \zeta^- = R$ by identifying $(x^\mu)$ with
$(x^\mu, (x^+ - R)/\sqrt{2}, (x^+ + R)/\sqrt{2}) \in {\cal N}$.  The \Mink\ %
space
is not closed for these conformal transformations, too.  Thus, the \Mink\ space
needs to be extended, resulting in the space $[{\cal N}] \cong S^1\times S^3$.


\section{The triality of null physics on conformal \Mink, \dS\, and \AdS\,
  spaces}

Let us now study the null physics on the conformal \Mink/\dS/\AdS\ spaces and
their relations via Weyl mappings. We also explain why there should be a
triality of the conjectured \AdS/\CFT-correspondence.

\noindent\subsection{The  triality of conformal \Mink, \dS\, and
\AdS\, spaces} \label{sect:triality}

According to the  above discussion,  \Mink\, and $H_\theta$ spaces and their
conformal extensions  with the same $R$  can be related by Weyl conformal
mappings as follows. A point in $dS$, say, is first viewed as a point in
$P_+\cap\mathcal{N}$.  Then a point in $P_-\cap\mathcal{N}$ equivalent to it
could be found, in general.  However, it is possible that a point in one space
could not be mapped into another space, or could not find an inverse image in
another space.  We solve it elsewhere.  Thus the mapping from the conformal
extension of $dS_4$ to that of $AdS_4$ is established.

For example, a point $\xi_+ := (\xi_+^0, \ldots, \xi_+^4) \in H_+$ can be
mapped to a point in $H_-$ with the following Beltrami coordinates:
\begin{equation}
  x_-^\mu := R\,\zeta^\mu/\zeta^5  = \xi_+^\mu.
  \label{dS2AdS}
\end{equation}
As another example, the Weyl conformal mapping sending a point with coordinates
$(x^\mu)$ in the \Mink\ space to a point in the \BdS\ space with  coordinates
$(x^\mu_+)$ reads
\begin{equation}
  x^\mu_+ = - \sqrt{2}\,x^\mu\,
    ({1 + \frac{1}{2R^2}\,\eta_{\rho\sigma}\,x^\rho x^\sigma})^{-1}.
  \label{M2dS}
\end{equation}
This is just the conformally flat coordinate transformation for the \BdS-metric
(\ref{bhl}), which is also known as a stereographic projection with an inverse
transformation
\begin{equation}
  x^\mu = - \sqrt{2}\,x_+^\mu ({1 \mp \sqrt{\sigma(x_+)}})^{-1}.
  \label{dS2M}
\end{equation}
The sign $\mp$ is opposite to the sign of $\xi^4\gtrless 0$ in the \BdS\ space.

It is important that the normal vectors of $P_+$, $P_-$ and $P_M$ are timelike,
spacelike and null, respectively, and that $P_+\cap\mathcal{N}$,
$P_-\cap\mathcal{N}$ and $P_M\cap\mathcal{N}$ are \dS, \AdS\ and \Mink\ space,
respectively.  This can be generalized: given a hyperplane off the origin, its
intersection with ${\cal N}$ is \dS, \AdS\ or \Mink\ if its normal vector is
timelike, spacelike or null, respectively.

We have shown that \Mink/\dS/\AdS\ spaces can all be conformally extended to
the same $[\mathcal{N}]$, so that they can be conformally mapped from one
to another.  A conformal transformation on one space is, in fact, also a
conformal transformation on another space.  And all these conformal
transformations are induced from the $SO(2,4)$ transformations, forming a group
$SO(2,4)/\mathbb{Z}_2$, due to the equivalence relation on $\mathcal{N}$.
Therefore, from the viewpoint  of conformal transformations, these three
kinds of spaces and the \CFT s on them are just same.  We refer to this fact as
the triality of conformal extensions of these spaces.

\subsection{Motion of free massless particles: Null geodesics}
\label{sect:nullGeod}

As was mentioned earlier, similar to a massive particle a free massless test
particle or a light signal in the \dS\ space has the  conserved 5-d angular
momentum (\ref{L5}). Namely,
\be
  L^{AB}:= \xi^A\,K^B - \xi^B\,K^A, \quad
  K^A := \frac{d\xi^A}{d\lambda},\quad
  \frac{d L^{AB}}{d\lambda} = 0,
  \nonumber \\ \label{K5}
\ee
where $\lambda=\lambda_+$ is an affine parameter.  The geometric meaning of
the Beltrami coordinates and the fact that geodesics have equations of straight
lines\cite{BdS, BdS2} imply that a geodesic is the intersection of $\Sigma$ and
\dS-hyperboloid $H_+$ in (\ref{Hpm}), where $\Sigma$ is some 2-d plane passing
through the origin of the 5-d Minkowski space $M_{1,4}$.  It can be proved that,
when the geodesic is null, it is in fact a straight line in $M_{1,4}$, having
the equation $\xi^A = \xi^A_0 + \lambda\,v^A$ for some constants $\xi^A_0$ and
$v^A$, satisfying $\eta_{AB} \xi^A_0 v^B = \eta_{AB} v^A v^B = 0$.  Thus the
5-d momentum $K^A = v^A$ of the null geodesic is also conserved:
\be
  \label{K^A}
  \frac{d K^A}{d\lambda}  = 0.
\ee

Using the relations~(\ref{Liexi+}),  we can obtain
\begin{equation}
  \label{Ls}
  L^{AB} = \frac{1}{\kappa^2}\,\frac{d\psi}{d\lambda}\,\mathcal{L}^{AB},
  \quad
  P^A = \frac{1}{\kappa^2 R^2}\,\frac{d\psi}{d\lambda}\,\mathcal{L}^{5A},
\end{equation}
where $\psi = \psi(\lambda)$ is certain a parameter and the 6-d angular
momentum $\mathcal{L}^{\hat{A}\hat{B}}$ is defined as
\begin{equation}
  {\cal L}^{\hat{A}\hat{B}}
  := \zeta^{\hat{A}}\,\frac{d\zeta^{\hat{B}}}{d\psi}
  - \zeta^{\hat{B}}\,\frac{d\zeta^{\hat{A}}}{d\psi}.
  \label{6-angmomentum}
\end{equation}
It is conserved if
\begin{equation}
  d\psi=\kappa^2 d\lambda.
  \label{psi-lambda}
\end{equation}
For a massless particle in the \AdS\ space, there are similar issues.

In the \Mink-case, the 4-d momentum $k_M^\mu$ and the  angular momentum
$l_M^{\mu\nu}$ are conserved for a light signal,
\begin{equation}
  k_M^\mu := \frac{dx^\mu}{d\lambda}, \quad
  l_M^{\mu\nu} := x^\mu k_M^\nu - x^\nu k_M^\mu.
\end{equation}
Similarly,  the 6-d angular momentum  defined by (\ref{6-angmomentum}) can be
related to $k^\mu_M$ and $l^{\mu\nu}_M$ by
\begin{eqnarray}
  && \mathcal{L}^{\mu\nu} =
  \frac{d\lambda}{d\psi}\,\kappa^2\,l_M^{\mu\nu}\,\quad
  \mathcal{L}^{4\nu}
  = \frac{1}{\sqrt{2}}\,(\mathcal{L}^{+\nu} - \mathcal{L}^{-\nu}),
  \\
  && \mathcal{L}^{5\nu}
  = \frac{1}{\sqrt{2}}\,(\mathcal{L}^{+\nu} + \mathcal{L}^{-\nu}),
  \quad
  \mathcal{L}^{45} = \mathcal{L}^{+-},
\end{eqnarray}
where
\be\nno\label{6-L}
  && \mathcal{L}^{-\nu} = \frac{d\lambda}{d\psi}\,\kappa^2 R\,k_M^\nu,\qquad\\
  && \mathcal{L}^{+\nu} = \frac{d\lambda}{d\psi}\,\kappa^2\,
    (x^+ k_M^\nu - x^\nu \frac{dx^+}{d\lambda} ), \\\nno
  && \mathcal{L}^{+-} = - \frac{d\lambda}{d\psi}\,\kappa^2\,R\,
    \frac{dx^+}{d\lambda}.\qquad
\ee
If eq.~(\ref{psi-lambda}) is satisfied, then the above 6-d angular momentum
is also conserved.

For a massless free particle, its equation of motion in $[{\cal N}]$ is not
unique in terms of $\zeta^{\hat{A}}$, because $\zeta^{\hat{A}}
= \zeta^{\hat{A}}(\psi)$ and
$
    \zeta^{\hat{A}} = \zeta'^{\hat{A}}(\psi')
    := \rho(\psi')\,\zeta^{\hat{A}} (\psi(\psi'))
$ are equivalent, with $\psi = \psi(\psi')$ a reparametrization.
Formally, there are the angular momenta
$\mathcal{L}^{\hat{A}\hat{B}}$ and
$\mathcal{L}'^{\hat{A}\hat{B}}(\psi')$ for the same particle.  But,
a reparametrization can always be chosen so that
$\mathcal{L}'^{\hat{A}\hat{B}}(\psi')$ is still conserved.

Consequently, the world line is lying in a 2-d plane $\Sigma$ passing through
the origin of $M_{2,4}$, which is also contained in \NsM\ except for the
origin.  Thus, the world line $\Sigma - \{0\}/\!\!\sim$ is a projective
straight line in $[\mathcal{N}]$: in the Beltrami coordinate on \dS\ or \AdS,
or in the \Mink\ coordinate, its equations look like
\begin{equation}
  x^\mu(s) = x^\mu_0 + s\,c^\mu,
  \label{eq:nullGeod}
\end{equation}
where $x^\mu_0$ and $c^\mu$ are some constants while $s$ is the
curve parameter. Hence, the world line is a null geodesic \cite{BdS,
BdS2}.  The relation of its 5-d angular momentum and ${\cal
L}^{\hat{A}\hat{B}}$ is as shown in eqs.~(\ref{Ls}), etc.  This
coincides with the well-known fact that null geodesics are
conformally invariant up to a reparametrization.

Thus, under the Weyl mapping (\ref{M2dS}) from $M_{1,3}$ to $\dS_4$, null
geodesics in the \Mink\ space $M_{1,3}$ is mapped to null geodesics in $\dS_4$.
What's more, for all these geodesics, no matter in $M_{1,3}$ or $\dS_4$,
their equations in the Minkowski coordinates or Beltrami coordinates are all
in the form of (\ref{eq:nullGeod}).  As indicated in section \ref{subsect:SRs},
all geodesics in \dS\ or \AdS\ spaces have the similar form of equations in any
Beltrami coordinate system.  Hence it is acceptable that Beltrami coordinates
in \dS\ or \AdS\ spaces play the role of Minkowski coordinates in SR.  That is,
Beltrami coordinates are inertial coordinates.

%
%

\subsection{On \CFT\ and  \AdS/\CFT\ correspondence} \label{sect:CFT}

Let us now consider other conformal issues on the conformal
\Mink/\dS/\AdS\ spaces and the relations among them.

The generators of the conformal group on the \Mink\ space are
\begin{eqnarray}
  && \hat{p}_\mu := \partial_\mu, \qquad
  \hat{l}_{\mu\nu} := x_\mu \, \partial_\nu - x_\nu\,\partial_\mu,
  \\
  && \hat{D} := x^\mu \, \partial_\mu,  \quad
  \hat{s}_\mu := - x\cdot x \,\partial_\mu + 2 x_\mu x^\nu\,\partial_\nu.
\end{eqnarray}
A \CFT\ in the \Mink\ space must be invariant under the action of these
generators.  The coordinates $x^\mu$ can be extended to be a set of coordinates
$(x^\mu, \kappa, \phi)$ on $M_{2,4} - \{\zeta^- = 0\}$, where $\kappa$ is the
scaling factor introduced before
\begin{eqnarray}
  \kappa = \frac{\zeta^-}{R}, \quad
  \phi := \eta_{\hat{A}\hat{B}}\,\zeta^{\hat{A}} \zeta^{\hat{B}}.
\end{eqnarray}
Thus the \Mink\ space is described by $\kappa = 1$ and $\phi = 0$.  Then it can
be verified that
\begin{eqnarray}
  && \hat{p}_\mu = \frac{1}{R}\,\hat{\mathcal{L}}_{+\mu}, \quad
  \hat{l}_{\mu\nu} = \hat{\mathcal{L}}_{\mu\nu},
  \label{Poincare} \\
  && \hat{D} = \hat{\mathcal{D}} + \hat{\mathcal{L}}_{-+}, \quad
  \hat{s}_\mu = 2x_\mu\,\hat{\mathcal{D}}
  + 2R\,\hat{\mathcal{L}}_{-\mu},
  \label{SK}
\end{eqnarray}
where
\begin{equation}
  \hat{\mathcal{D}} := \zeta^{\hat{A}}\,
    \frac{\partial}{\partial\zeta^{\hat{A}}}
\end{equation}
is the generator of scaling in $M_{2,4}$, while
\begin{equation}\label{gso24}
  \hat {\cal L}_{\hat A \hat B}
  := \zeta_{\hat A} \, \frac{\partial}{\partial \zeta^{\hat B}}
  - \zeta_{\hat B} \, \frac{\partial}{\partial\zeta^{\hat A}}
\end{equation}
are generators of  \cg.  Since $\hat{\mathcal{D}}$ is commutative with each
$\hat{\mathcal{L}}_{\hat{A}\hat{B}}$, it does not matter that the conformal
generators of the \Mink\ space differ from those of $M_{2,4}$ by a vector field
along $\hat{\mathcal{D}}$ [see, eqs.~(\ref{SK})].  This coincides with (i) the
idea that the equivalence relation $\sim$ will be considered on $\mathcal{N}$,
and (ii) the fact the conformal transformations in the \Mink\ space is induced
from, but not the same as, the \CG-transformations on $\mathcal{N}$.  In fact,
a quantity in the \Mink\ space can be realized by homogeneous function of degree
zero in $M_{2,4} - \{0\}$.  In this way $\hat{\mathcal{D}}$ somehow could be
dropped directly.

Generators of conformal transformations on \dS/\AdS\ spaces, or specially on
\BdS/\BAdS\ spaces, can also be given as the ones of \cg.  Thus, they can be
related by the Weyl conformal mappings such as (\ref{dS2AdS}) and (\ref{M2dS}).
Correspondingly, the \CFT s in these spaces are also related by these mappings.
Since the Maxwell equations are the simplest \CFT, as an illustration, we show
how the sourceless Maxwell equations
\begin{equation}
  d \vect{F} = 0, \qquad *\  d *\vect{F} = 0,
  \label{Maxwell}
\end{equation}
where
$*$ is the Hodge dual operator, are related among them.

Consider  the  Weyl conformal mapping $\psi: M_{1,3} \rightarrow \dS_4$ as
shown in eq.~(\ref{M2dS}):
\begin{equation}
  \psi^* \vect{g} = \Omega^2\,\vect{\eta}, \qquad
  \Omega = \sqrt{2} \, \Big( 1 - \frac{1}{2R^2}\,\eta_{\mu\nu}\,x^\mu x^\nu
    \Big)^{-1},
\end{equation}
with  $\vect{g}$ the metric (\ref{bhl}) of $\BdS_4$,  $\vect{\eta}$ the one in
(\ref{ihN}).  If $\vect{F}_{\dS}$ is the Maxwell field in $\dS_4$, its
equations follow
\begin{equation}
  d \vect{F}_{\dS} = 0, \qquad
  \star\ d \star\vect{F}_{\dS} = 0,
  \label{Maxwell-dS}
\end{equation}
where $\star$ is the dual operator with respect to $\vect{g}$.  We pull
$\vect{F}_{\dS}$ back to the \Mink\ space, resulting in
\begin{equation}
  \vect{F} = \psi^* \vect{F}_{\dS}.
  \label{eq:F}
\end{equation}
Thus $d\vect{F} = d \, (\psi^*\vect{F}_{\dS}) = \psi^* d\vect{F}_{\dS} = 0$ is
satisfied. It can be verified that
\begin{eqnarray*}
  \psi^*(\star\,d\star\vect{F}_{\dS})
  = \Omega^{-2}\,[*\,d * \vect{F}].
\end{eqnarray*}
Therefore, in the \Mink\ space, $\vect{F}$ as in eq.~(\ref{eq:F}) is a
sourceless electromagnetic field: eqs.~(\ref{Maxwell}) are satisfied.
In this way the Weyl conformal mapping $\psi: M_{1,3} \rightarrow \BdS_4$
relates a sourceless electromagnetic field $\vect{F}_{\dS}$ in the \BdS\ space
to a sourceless $\vect{F}$ in the \Mink\ space.

Similarly, the approach can be applied to other \CFT s between \dS\ and
\AdS\ spaces, \AdS\ and \Mink\ spaces and so on.  Basically, the \CFT s in
\Mink/\dS/\AdS\ spaces, in which all the relevant fields are assumed to behave
well as the infinity points are approached, can be unified together.  The
former is merely a realization of the latter.

For the \AdS/\CFT\ correspondence, there should also be a triality.

The $\AdS_5$ can be embedded into $M_{2,4}$ as a hypersurface $\mathcal{S}$:
\be
  {\cal S}:~ \eta_{\hat{A}\hat{B}}\,\zeta^{\hat{A}} \zeta^{\hat{B}} = R_5^2,
  \quad {\rm  with}~~R_5 > 0.
\ee If antipodal points in $\mathcal{S}$ are identified, the
resulted space, denoted by $\mathcal{S}/\mathbb{Z}_2$, is still {
homeomorphic} to $\mathcal{S}\cong \AdS_5$.  In the projective space
$\mathbb{R}P^5 = M_{2,4} - \{0\}/\sim$, the quotient space of those
$\zeta^{\hat{A}}$ satisfying $\eta_{\hat{A}\hat{B}}\,
\zeta^{\hat{A}} \zeta^{\hat{B}} > 0$ are  homeomorphic to
$\mathcal{S}/\mathbb{Z}_2 \cong \AdS_5$.  Identifying $\AdS_5$ with
this quotient space, then
\begin{equation}\label{boundary-ads5}
  \partial(\AdS_5) = [\mathcal{N}].
\end{equation}

Thus, due to the triality of the \CFT s in conformal \Mink/\dS/\AdS\ spaces,
there should be three {\AdS/\CFT} correspondences starting from the well-known
\AdS/\CFT\ correspondence \cite{adscft}. Namely, there should be  the
{\AdS/\CFT} correspondence between $AdS_5$ and the $\dS_4$/$\AdS_4$,
respectively, in addition to that between $\AdS_5$ and the \Mink\ space.
Clearly, this triplet of the \AdS/\CFT\ correspondence can be generalized to
any dimensions whenever the \AdS/\CFT\ correspondence is conjectured.

\section{Concluding remarks}

It is emphasized that since there are three kinds of special relativity with
$ISO(1,3)/SO(1,4)/SO(2,3)$ invariance, respectively, there should be the
triality of their conformal extensions.  Actually, similar to the {\it inertia}
feature of the \Mink\ coordinates on the \Mink\ space \M, the Beltrami systems
on the \dS/\AdS\ space $\dS_4$/$\AdS_4$ play the same role of {\it inertia},
respectively. In each of three kinds of special relativity, massive particles
or massless particles move uniformly along the timelike or null geodesics with
constant coordinate velocity components, respectively.  Thus, their motions are
of {\it inertia}.  In fact, as was mentioned, three kinds of special relativity
with the Poincar\'e/\dS/\AdS-invariance can first be set up based on the
principle of relativity and the postulate on the invariant universal
constant(s). While for massless particles and light signals, the symmetry
should be enlarged from the Poincar\'e/\dS/\AdS-group to the conformal group
$SO(2,4)/\mathbb{Z}_2$, respectively. On the other hand, we may also set up
these realizations with the conformal group invariance, basing on the principle
of relativity with respect to  massless particles and light signals alone, in
principle.  And for the massive cases, the theory with conformal group symmetry
should break back to the special relativity of Poincar\'e/\dS/\AdS\ invariance,
respectively.

We have shown that  the conformal extensions of three kinds of special
relativity can be realized on {$[{\cal N}]$,} the projective null cone modulo
projective equivalence, as the 4-d $SO(2,4)/\mathbb{Z}_2$ conformal
\Mink/\dS/\AdS\ spaces with  the same constant $R$.  In general, each space may
have its own constant $R$ while our approach works similarly. The Weyl conformal
mappings among these spaces,  the motions of light signals and the \CFT s
on them have been given.  Thus, there should be a triality for these conformal
issues.  Since $\partial(\AdS_5)=[{\cal N}]$, there should be a triplet of
the conjectured \AdS/\CFT\ correspondence.

It should also be noted that the supersymmetry extension of the \cg\ can also
be viewed as that of \dS/\AdS-algebras as subalgebras in bosonic sector, in
addition to the conventional approach with Poincar\'e algebra as a subalgebra.


\begin{acknowledgments}

We would like to thank Z.~Chang, C.-G.~Huang, W.L.~Huang, Q.K.~Lu,
X.C.~Song, S.K.~Wang, K.~Wu, X.N.~Wu and C.J.~Zhu for valuable
discussions. C.-G.~Huang attended partly and made contributions to
this work. This work is partly supported by NSFC under Grant
Nos.~10375087, 90503002, 10347148, 10547002, 10505004
and 10605005.
\end{acknowledgments}

\end{document}